\documentclass[aps,preprint,nofootinbib,superscriptaddress]{revtex4}

\usepackage{amssymb}

\usepackage{epsfig}
\usepackage[bookmarksnumbered,bookmarksopen,colorlinks,citecolor=blue,linkcolor=blue]{hyperref}

\begin{document}

\title{ Improvement on Fermionic properties and new isotope production in molecular dynamics simulations }

\author{Ning Wang}
\email{wangning@gxnu.edu.cn}\affiliation{ Department of Physics,
Guangxi Normal University, Guilin 541004, People's Republic of
China }
\affiliation{ State Key Laboratory of Theoretical Physics, Institute of Theoretical Physics, Chinese Academy of Sciences, Beijing 100190, People's Republic of China}

\author{Tong Wu}
\affiliation{ Department of Physics,
Guangxi Normal University, Guilin 541004, People's Republic of
China }

\author{Jie Zeng}
\affiliation{ Department of Physics,
Guangxi Normal University, Guilin 541004, People's Republic of
China }

\author{Yongxu Yang}
\affiliation{ Department of Physics,
Guangxi Normal University, Guilin 541004, People's Republic of
China }

\author{Li Ou}
\affiliation{ Department of Physics,
Guangxi Normal University, Guilin 541004, People's Republic of
China }
\affiliation{ State Key Laboratory of Theoretical Physics, Institute of Theoretical Physics, Chinese Academy of Sciences, Beijing 100190, People's Republic of China}

\begin{abstract}

  By considering momentum transfer in the Fermi constraint procedure, the stability of the initial nuclei and fragments produced in heavy-ion collisions can be further improved in the quantum molecular dynamics simulations. The case of the phase space occupation probability larger than one is effectively reduced with the proposed procedure. Simultaneously, the energy conservation can be better described for both individual nuclei and heavy-ion reactions. With the revised version of the improved quantum molecular dynamics (ImQMD) model, the fusion excitation functions of $^{16}$O+$^{186}$W and the central collisions of Au+Au at 35 AMeV are re-examined. The fusion cross sections at sub-barrier energies and the charge distribution of fragments are relatively better reproduced due to the reduction of spurious nucleon emission. The charge and isotope distribution of fragments in Xe+Sn, U+U and Zr+Sn at intermediate energies are also predicted. More unmeasured extremely neutron-rich fragments with $Z=16-28$ are observed in the central collisions of $^{238}$U+$^{238}$U than that of $^{96}$Zr+$^{124}$Sn, which indicates that multi-fragmentation of U+U may offer a fruitful pathway to new neutron-rich isotopes.

\end{abstract}
\maketitle

\begin{center}
\textbf{I. INTRODUCTION}
\end{center}

The microscopic dynamics simulations of heavy-ion collisions (HICs) are important not only for the study of multi-fragmentation process at intermediate energies \cite{More93} but also for the study of fusion reactions at energies near Coulomb barrier. The nuclear multi-fragmentation process can be used to investigate the liquid-gas phase transition \cite{Bord08,Mull95}, the mechanisms of fragment production in highly excited
nuclear systems \cite{Aich88,AuAu,Ono92}, and as well as the equation of state (EOS) in asymmetric nuclear matter, e.g. nuclear symmetry energy \cite{Bar02,Tsang01,Tsang09}. In addition, the measurement for the reaction of 25 AMeV $^{86}$Kr+ $^{64}$Ni \cite{Sou02,Sou03} and the microscopic dynamics simulations for Kr+Ni at energies well above the Coulomb barrier but below the Fermi energy \cite{Fount14} indicates that this kind of reaction at this energy regime may offer a fruitful pathway to extremely neutron-rich nuclei, towards the neutron-drip line. It is known that the masses of a large number of extremely neutron-rich nuclei such as three magic nuclei $^{46}$Si, $^{60}$Ca, $^{78}$Ni predicted by the Weizs\"acker-Skyrme (WS4) mass model \cite{WS4,WS4a} have not yet been measured, due to the difficulties in producing these nuclei and their short half-lives. These extremely neutron-rich nuclei play an important role for testing nuclear mass models and for exploring nuclear symmetry energy. The synthesis of very neutron-rich nuclides through multi-fragmentation or deep inelastic transfer reactions are therefore of exceptional importance to advance our understanding of nuclear structure at the extreme isospin limit of the nuclear landscape.

Although the compression process and the multifragment de-excitation of hot compressed composite nuclei formed in heavy-ion collisions at intermediate incident energies can be reasonably well described by the statistical multi-fragmentation (SMM) model \cite{Bond95,Bot01,Buy05} and some transport models such as the quantum molecular dynamics (QMD) model together with a statistical decay model \cite{QMD,John97,Hag94}, the uncertainty of model predictions especially for the primary fragments from different transport models and the time to combining the statistical decay model are still large. The stability of the initial nuclei and of primary fragments produced in HICs should be further tested. As a semi-classical microscopic dynamics model, the QMD model was proposed for simulating heavy-ion collisions at intermediate and high energies. In the QMD model, no anti-symmetrization is carried out and its effects are usually simulated by using a phenomenological Pauli potential \cite{Pei92} and the collision term. To cure the problem of Fermions in phase space, Reinhard and Suraud proposed a modification of the Vlasov equation which guarantees that the semiclassical dynamics obeys the Pauli principle and relaxes towards a Fermi equilibrium, by introducing a dedicated collision term \cite{Rein96}. For the purpose of simulating the Pauli exclusion principle, a momentum-dependent two-body repulsion was introduced in Refs.\cite{Wilets,Dorso87,Dorso89}. Although the proposed Pauli potentials and collision term met with some success, the stability of the model especially for describing a bound nuclear system after a long time evolution should be further improved. The spurious nucleon emission in heavy-ion reactions or in an individual heavy bound nucleus becomes serious in the traditional QMD model after evolution of several hundreds fm/c, which causes uncertainties in the description of the reaction yields and cross sections.

With great efforts to develop the QMD model, some different extended versions of the QMD model such as IQMD \cite{IQMD93,IQMD99}, CoMD \cite{constrain,Maru02,Sou14,Erg15}, ImQMD \cite{ImQMD2002,ImQMD2004,ImQMD2006,ImQMD2010}, EQMD \cite{Maru96,Ma14} and UrQMD \cite{UrQMD,LiQF} have been proposed in the literature. To extend the QMD model for the study of heavy-ion reactions at energies around the Coulomb barrier, the ImQMD model was proposed in which the standard Skyrme force is adopted for describing not only the bulk properties but also the surface properties of nuclei, and the phase-space occupation constraint method is used following the CoMD model to simulate the effects of anti-symmetrization and to improve the stability of an individual nucleus. In this constraint (also called Fermi constraint), the phase space occupation probability $\bar f_{i}$ of the $i$-th particle is checked during the propagation of nucleons. If $\bar f_{i}>1$, the momentum of the particle $i$ is randomly changed by a series of two-body "elastic scattering" between this particle and its neighboring particles which is similar to the procedure in the traditional collision term of QMD simulations but neglecting the influence of nucleon-nucleon ($NN$) scattering cross sections. The Pauli blocking condition is simultaneously checked after the momentum redistribution via the two-body "elastic scattering". For each $NN$ collision we evaluate the occupation probability after the "elastic scattering". If the occupation probabilities are both less than one, the collision is accepted, and rejected otherwise. With these modifications in the ImQMD model, the stability of the initial nuclei and fragments is significantly improved \cite{ImQMD2014}. However, there is still a few spurious nucleon emissions for heavy bound systems after time evolution of a few thousands fm/c and the total energy of the system slightly increases due to the frequently abrupt change of particle momentum in the two-body "elastic scattering". If neglecting the Fermi constraint, the time evolution by classical equations of motion in the QMD simulations surely breaks the initial Fermi-Dirac distribution which evolves into a classical Boltzmann one \cite{constrain}. The Fermi constraint adopted in the previous version of ImQMD model mainly affects the low momentum part of the momentum distribution. The long tail (high momentum part) of the momentum distribution which is much longer than that of the Fermi-Dirac distribution at low temperature can not be effectively improved since the phase space occupation probability of the particle with high momentum and low density is generally smaller than one and cannot be constrained with the procedure. It is therefore necessary to further improve the Fermi constraint for a better description of the stability and energy conservation.

To further improve the stability of the initial nuclei and fragments, the Fermi constraint procedure is modified in the version ImQMD-v2.2 \cite{Wang15} and the momentum transfer (similar to the two-body inelastic scatter but without new particle production) in the momentum re-distribution process is simultaneously considered in addition to the "elastic scattering" involved in v2.1. According to the Fermi-Dirac distribution at low temperature, the momenta of any two neighboring nucleons in a nucleus or fragment should not deviate from each other very large. If the difference between the momentum of a nucleon and that of its any neighboring nucleons is larger than Fermi momentum, $|\vec{p}_i-\vec{p}_j|>p_F$, $i=1,2...N, j\neq i$, with the Fermi momentum $p_F=260$ MeV/c, a tiny part of momentum $\vec{p} f_t $ of the nucleon with a higher momentum will be transferred to the other one. For heavy-ion fusion reactions, the transfer factor is set as $f_t=5\times10^{-6}$ for the parameter set IQ3a \cite{ImQMD2014} which guarantees that the total momentum and energy of the system are well conserved in the simulations. It was found that the consideration of the momentum transfer in the Fermi constraint can significantly reduce the number of spurious emitted nucleons. It is interesting to investigate the influence of this kind of momentum transfer on properties of nucleons and density evolution of reaction system.

In this work, we further check the Fermionic properties of nucleons and energy conservation in the nuclear system during the ImQMD simulations. Simultaneously, we systematically investigate the nuclear multi-fragmentation and new isotope production with the ImQMD-v2.2 model. The structure of this paper is as follows: In sec. II, the mean-field part of the ImQMD model will be introduced. In sec. III, the fusion reaction $^{16}$O+$^{186}$W and the multi-fragmentation of $^{197}$Au+$^{197}$Au at an incident energy of 35 AMeV will be re-examined, and the energy conservation will be checked. In sec. IV, we investigate the charge and isotope distribution of fragments in Xe+Sn, U+U and Zr+Sn at intermediate incident energies. Finally a brief summary is given in Sec. IV.

\begin{center}
\noindent{\bf {II. MEAN-FIELD IN ImQMD }}\\
\end{center}

In the improved quantum molecular dynamics simulations, both the self-consistently generated mean-field and the momentum re-distribution in the Fermi constraint affect the movements of nucleons. For the mean-field part, each
nucleon is represented by a coherent state of a Gaussian wave
packet. The density distribution function $\rho$ of a system reads
\begin{equation} \label{1}
\rho(\mathbf{r})=\sum_i{\frac{1}{(2\pi \sigma_r^2)^{3/2}}\exp
\left [-\frac{(\mathbf{r}-\mathbf{r}_i)^2}{2\sigma_r^2} \right ]},
\end{equation}
where $\sigma_r=\sigma_0+\sigma_1 A^{1/3}$ represents the spatial spread of the wave packet \cite{Li2013}.
The propagation of nucleons is governed by the mean field,
\begin{equation} \label{2}
\mathbf{\dot{r}}_i=\frac{\partial H}{\partial \mathbf{p}_i}, \; \;
\mathbf{\dot{p}}_i=-\frac{\partial H}{\partial \mathbf{r}_i},
\end{equation}
where $r_i$ and $p_i$ are the center of the $i$-th wave packet in
the coordinate and momentum space, respectively. Euler algorithm is adopted to compute new positions and momenta at time $t+ \Delta t$. The time step in the ImQMD calculations is set as $\Delta t=1$ fm/c. The Hamiltonian
$H$ consists of the kinetic energy
$T=\sum\limits_{i}\frac{\mathbf{p}_{i}^{2}}{2m}$ and the effective
interaction potential energy $U$ which is written as the sum of
the nuclear interaction potential energy $U_{\rm
loc}=\int{V_{\rm loc}(\textbf{r})d\textbf{r}}$ and of the Coulomb
interaction potential energy. Where $V_{\rm loc}(r)$ is the potential energy density that is
obtained from the effective Skyrme interaction, in which the spin-orbit term is not involved:
\begin{equation}
V_{\rm
loc}=\frac{\alpha}{2}\frac{\rho^2}{\rho_0}+\frac{\beta}{\gamma+1}\frac{\rho^{\gamma+1}}{\rho_0^{\gamma}}+\frac{g_{\rm
sur}}{2\rho_0}(\nabla\rho)^2
+g_{\tau}\frac{\rho^{\eta+1}}{\rho_0^{\eta}}+\frac{C_s}{2\rho_0}[\rho^2-k_s(\nabla\rho)^2]\delta^2
\end{equation}
where $\delta=(\rho_n -\rho_p)/(\rho_n +\rho_p)$ is the isospin
asymmetry. In Table I we list the model parameters IQ3a \cite{ImQMD2014}  adopted in the calculations. The corresponding value of the incompressibility coefficient of symmetric nuclear matter is about 225 MeV.

\begin{table}
 \caption{ Parameter set IQ3a \cite{ImQMD2014}.}
\begin{tabular}{lccccccccccc}
\hline Parameter & $\alpha $ & $\beta $ & $\gamma $ &$%
g_{\rm sur}$ & $ g_{\tau }$ & $\eta $ & $C_{s}$ & $\kappa _{s}$ &
$\rho
_{0}$ & ~~$\sigma_0$~~ & ~~$\sigma_1$~~ \\
 & (MeV) & (MeV) &  & (MeVfm$^{2}$) & (MeV) &  & (MeV) & (fm$^{2}$) &
 (fm$^{-3}$) & (fm) & (fm) \\ \hline
IQ3a & $-207$ & 138 & 7/6 & 16.5 & 14 & 5/3 &  34  & 0.4  & 0.165 & 0.94 & 0.02\\
  \hline
\end{tabular}
\end{table}

In the present calculations, the traditional collision term in the QMD simulations (in which the free or in-medium nucleon-nucleon scattering cross sections are involved) is switched off and the initialization of the reaction partners is as follows: to obtain the reasonable initial nuclei, the nucleon positions are sampled within two hard spheres in which the neutron-skin thickness is simultaneously considered. With the sampled nucleon positions, the nuclear potential energy of the nucleus can be calculated. The momentum of the $i$-th nucleon is then sampled within the local Fermi sphere with a radius $ \hbar [3\pi^2 \rho_q ({\bf r}_i) ]^{1/3}-w_p$, where $q=n$ for neutrons and $q=p$ for protons. $w_p$ is to consider the influence of the width of the wave-packet in the momentum space and its value is determined by the experimental binding energy $BE$ (in negative value) of the sampled nuclei. In this work, if the ground state energy of the nucleus calculated falls into the range of $BE \pm 0.05$ MeV and simultaneously the distance between any two identical nucleons in the phase space fulfills the uncertainty relation $\frac{4\pi}{3}r_{ij}^3\cdot \frac{4\pi}{3}p_{ij}^3 \ge \frac{h^3}{8}$ \cite{Zhang06, Ou08} (where $h$ is Planck's constant, $r_{ij}$ and $p_{ij}$ are the distances between the centers of the wave packets of two nucleons $i$ and $j$ in coordinate and
momentum space), the sampled nucleus will be used in the ImQMD simulations.

\begin{center}
\textbf{III. COMPARISON OF THE CALCULATED RESULTS}
\end{center}

\begin{figure}
\includegraphics[angle=0,width=0.65\textwidth]{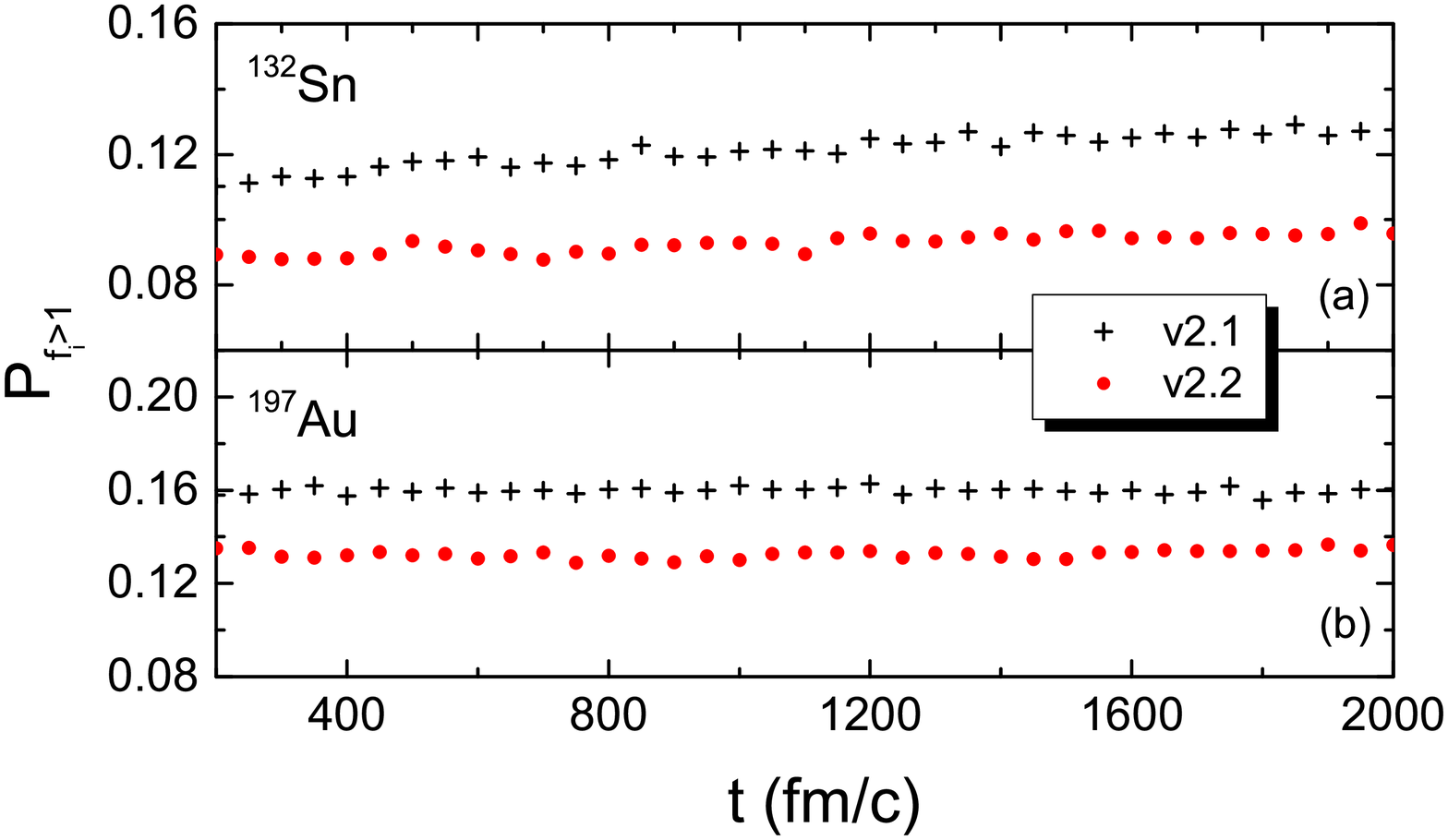}
\caption{(Color online) Probability of nucleons with $f_i>1$ in the individual nuclei as a function of time. The crosses and circles denote the results with version v2.1 and v2.2, respectively.}
\end{figure}

\begin{figure}
\includegraphics[angle=0,width=0.7 \textwidth]{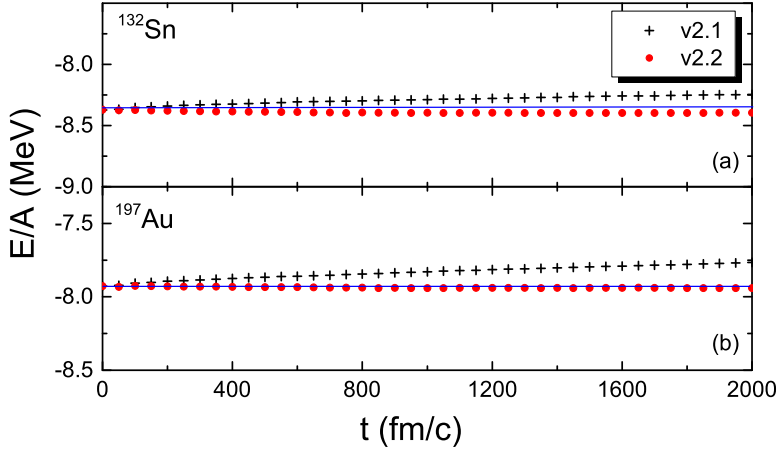}
\caption{(Color online) Time evolution of the binding energies of $^{132}$Sn and $^{197}$Au. The crosses and circles denote the results with version v2.1 and v2.2, respectively. The blue lines denote the corresponding measured binding energies per particle for the two nuclei.   }
\end{figure}

In Ref. \cite{Wang15}, it was found that considering momentum transfer in the Fermi constraint procedure, the average numbers of spurious emitted nucleons in the initial nuclei at $t=2000$ fm/c are reduced by $50\%$ (to 0.56) for $^{92}$Zr and $33\%$ (to 1.75) for $^{132}$Sn, respectively. To understand the reason for the reduction of the spurious emission, we investigate the density distribution, momentum distribution and phase space occupation probability of nucleons in an individual nucleus. We find that the new Fermi constraint procedure does not significantly affect the density and momentum distribution of the individual bound nuclei. However, the phase space occupation probabilities $\bar f_{i}$ of nucleons are evidently influenced, and the number of "pseudo" nucleons (with $\bar f_{i}>1$) is effectively suppressed. In this work, we check the probability $P_{f_i>1}$ of "pseudo" nucleons in the individual nuclei. In Fig. 1, we show the time evolution of the probability $P_{f_i>1}$ in $^{132}$Sn and $^{197}$Au. The circles and crosses denote the results with and without the momentum transfer via "inelastic scattering" being considered, respectively. One can see from the figure that there are about 12$\%$ "pseudo" nucleons in the individual $^{132}$Sn and 16$\%$ in $^{197}$Au, if only the two-body "elastic scattering" being taken into account in the Fermi constraint (v2.1). When the momentum transfer via "inelastic scattering" is considered simultaneously, the number of "pseudo" nucleons is evidently reduced in both $^{132}$Sn and $^{197}$Au. It seems that the momentum re-distribution improves the phase space occupation probability, which is helpful to further reduce the spurious emission of nucleons in the bound nuclei. The time evolution of the binding energy of the individual nuclei is also checked. In Fig. 2, we compare the time evolution of the binding energy per particle of $^{132}$Sn and $^{197}$Au. We find that energy conservation of bound nuclei can be obviously improved with the new version (v2.2) of ImQMD, by considering the momentum transfer and introducing a transfer factor $f_t$ with a fixed value of $5\times10^{-6}$ in the calculations.

\begin{figure}
\includegraphics[angle=0,width=0.7 \textwidth]{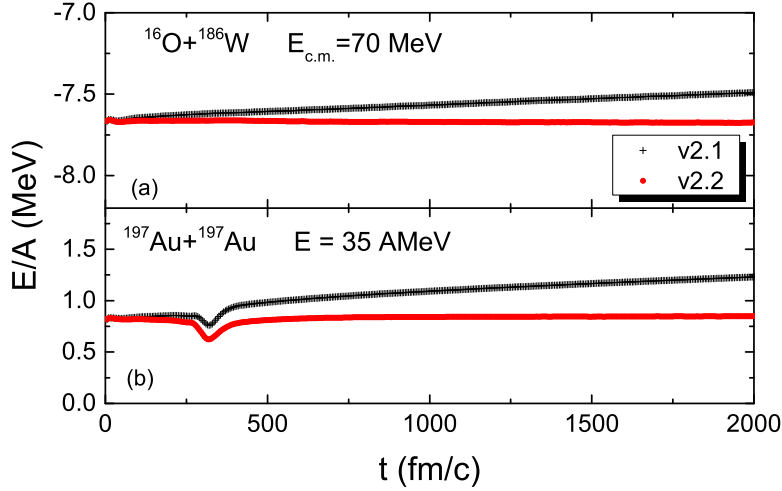}
\caption{(Color online)  The same as Fig. 2, but for fusion reaction of $^{16}$O+$^{186}$W and central collision of $^{197}$Au+$^{197}$Au at an incident energy of 35 AMeV.  }
\end{figure}

\begin{figure}
\includegraphics[angle=0,width=0.7 \textwidth]{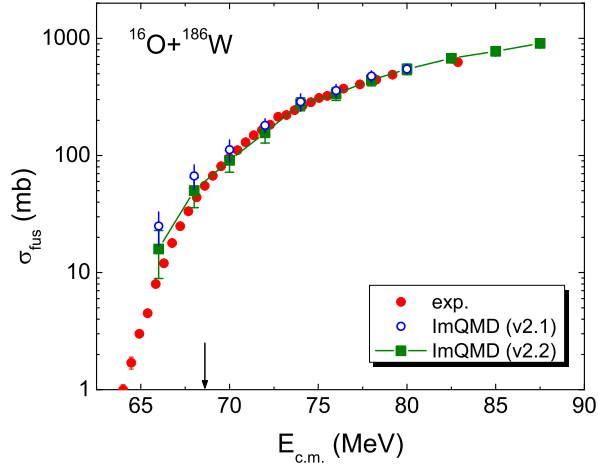}
\caption{(Color online)Fusion excitation functions of $^{16}$O+$^{186}$W. The solid circles denote the experimental data taken from Refs. \cite{OSm}. The solid squares and open circles denote the results of ImQMD with version v2.2 and v2.1, respectively. The statistical errors in the ImQMD calculations are given by the error bars. The arrow denotes the position of the most probable barrier height.}
\end{figure}

With the ImQMD-v2.2 model, the fusion reaction of $^{16}$O+$^{186}$W and the central collisions of $^{197}$Au+$^{197}$Au at an incident energy of 35 AMeV are re-examined. In Fig. 3, we show the time evolution of the energy of the total reaction system. Similar to the cases for the individual nuclei, the energy conservation can be better described comparing with the results from the version v2.1. The fluctuation of the binding energy at about 300 fm/c in Au+Au is due to the violent compression process in which two-body "scattering" strongly affects the propagation of nucleons at the densities much higher than the normal density. These calculations indicate that the modification of the Fermi constraint not only improves the phase space occupation probability of nucleons, but also better describes the time evolution of system energy. In addition, the fusion excitation function of $^{16}$O+$^{186}$W is calculated with the two versions of the ImQMD model. The calculation results are shown in Fig. 4. One sees that at energies below the Coulomb barrier, the results from v2.2 are lower than those from v2.1 due to the fewer spurious emission in the simulations of v2.2.

\begin{figure}
\includegraphics[angle=0,width=0.8\textwidth]{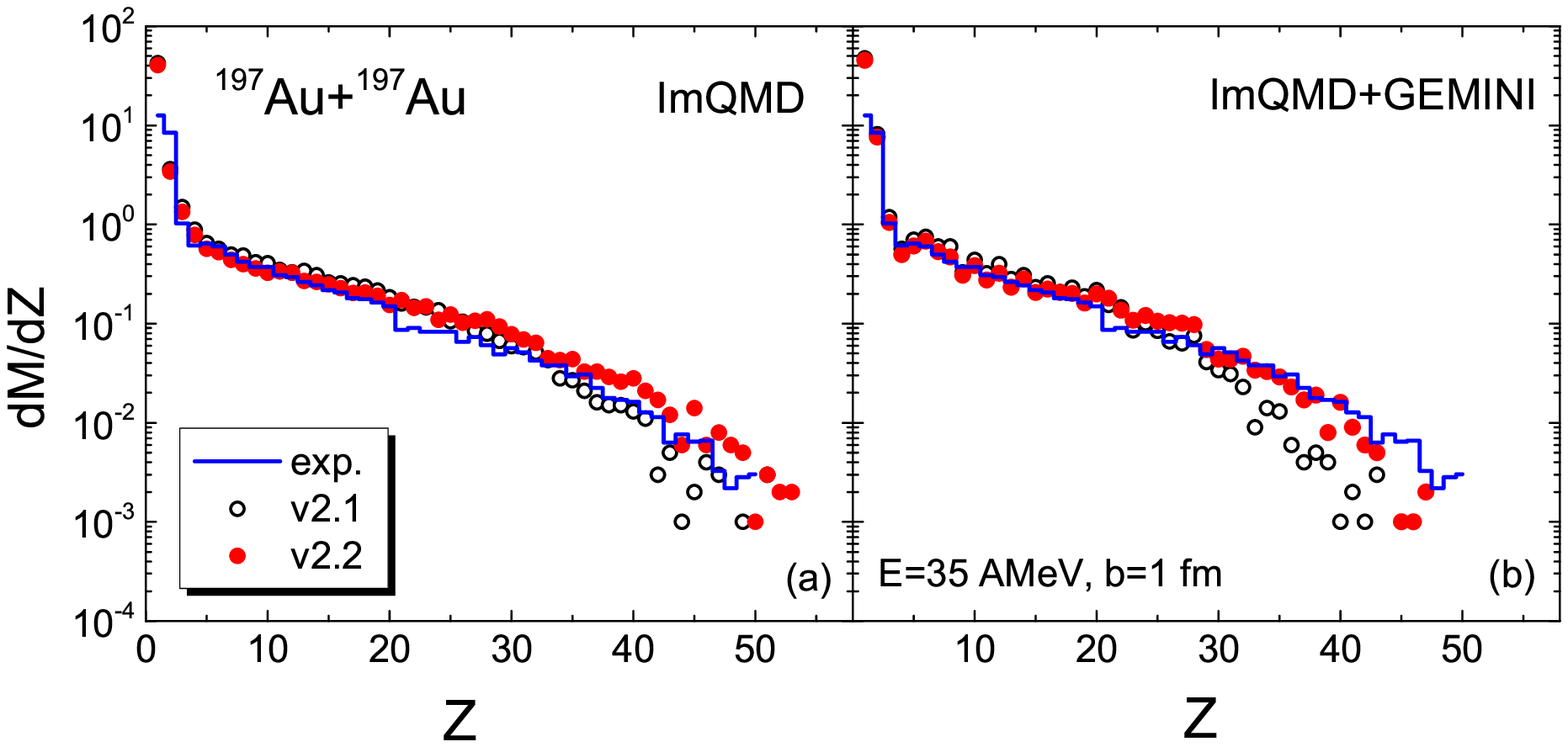}
\caption{(Color online) Charge distribution of fragments in $^{197}$Au+$^{197}$Au at an incident energy of 35 AMeV and $b=1$ fm. (a) and (b) denote the results of ImQMD simulations at $t=2000$ fm/c without and with the statistical model (GEMINI) being combined, respectively.     }
\end{figure}

\begin{figure}
\includegraphics[angle=0,width=0.75 \textwidth]{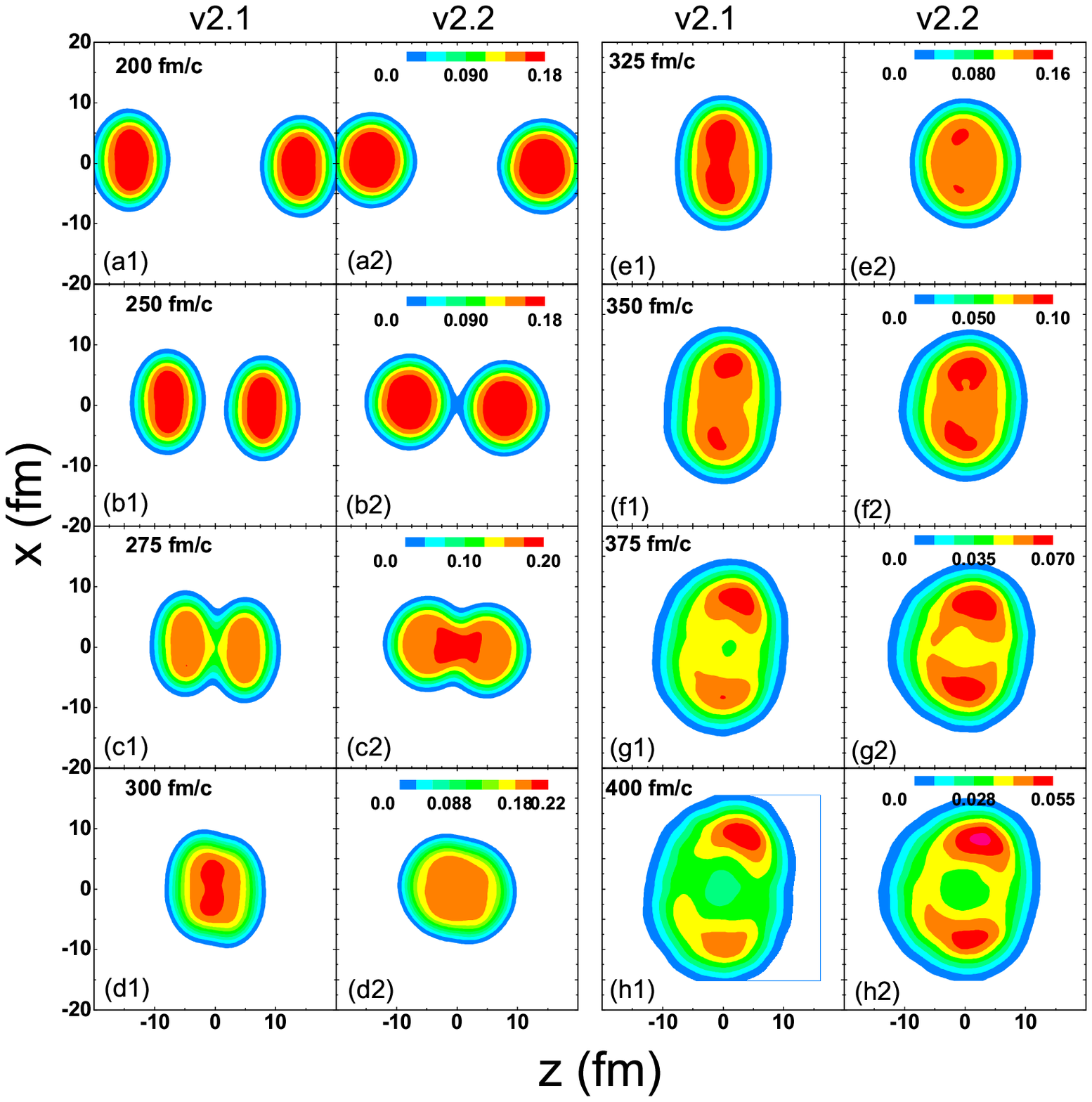}
\caption{(Color online) Time evolution of nuclear density in $^{197}$Au+$^{197}$Au at an incident energy of 35 AMeV and $b=1$ fm. }
\end{figure}

In Fig. 5, we show the charge distribution of fragments in central collisions of $^{197}$Au+$^{197}$Au at an incident energy of 35 MeV per particle and $t=2000$ fm/c. The circles in Fig. 5(a) denote the distributions of the primary fragments directly from the ImQMD simulations. The curve denotes the experimental data \cite{AuAu}. One sees that for light and intermediate fragments, the results from the two versions are close to each other very much. The yields of heavy fragments from v2.2 are larger than those from v2.1 due to the reduction of spurious nucleon emission. The experimental data are reproduced reasonably well. To further investigate the influence of the secondary decay of the excited primary fragments, the statistical model (GEMINI code with default values \cite{Char88,Char98}) is combined after the ImQMD simulations. The corresponding results are presented in Fig. 5(b). One sees that the experimental data are relatively better reproduced with the version v2.2. We also note that the results at $t=3000$ fm/c are close to those at $t=2000$ fm/c. To further see the influence of the new Fermi constraint on the compression-expansion process, we investigate the time evolution of density distribution in the HICs of Au+Au. In Fig. 6, we compare the density distribution of Au+Au by adopting the two versions of ImQMD. We note that the shapes of the reaction partners before touching configuration are more spherical in v2.2 comparing with those in v2.1. At $t=275$ fm/c, the neck density is much higher in v2.2. In the whole compression-expansion process, the elongation of reaction system along x-axis is more evident in v2.1. It seems that the momentum re-distribution via the two-body "inelastic scattering" could also influence the angular distribution of fragments in the HICs.

\begin{center}
\textbf{IV. PRODUCTION OF NEW ISOTOPES IN MULTI-FRAGMENTATION}
\end{center}

With the ImQMD-v2.2 model, we also study the production of new neutron-rich nuclides in multi-fragmentation process of heavy-ion collisions.  We first investigate the charge and isotope distribution of fragments in $^{129}$Xe+$^{120}$Sn at incident energy of 50 AMeV, in which the experimental data for the charge distribution are available. Here, we focus our investigations on the primary fragments produced at the central collisions, since the influence of the secondary decay (e.g. fission of fragments) on the charge distribution for the intermediate mass fragments might be negligible at such an energy region which can also be observed from Fig. 5.

\begin{figure}
\includegraphics[angle=0,width=0.85 \textwidth]{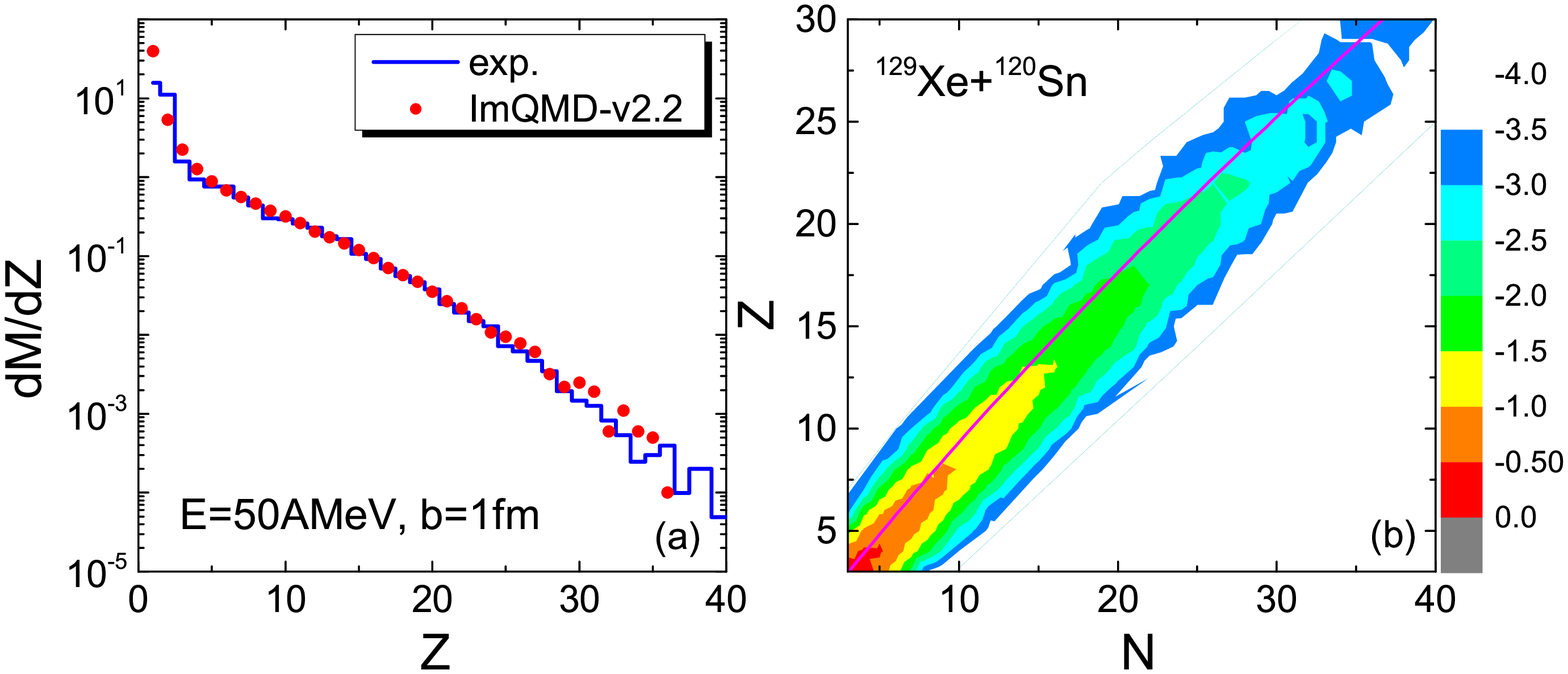}
\caption{(Color online) (a) Charge and (b) isotope distribution of fragments in  $^{129}$Xe+$^{120}$Sn at an incident energy of 50 AMeV and $b=1$ fm. The experimental data in (a) are taken from Ref.\cite{Taba}. The curve in (b) denotes the $\beta$-stability line described by Green's formula. }
\end{figure}

\begin{figure}
\includegraphics[angle=0,width=0.85  \textwidth]{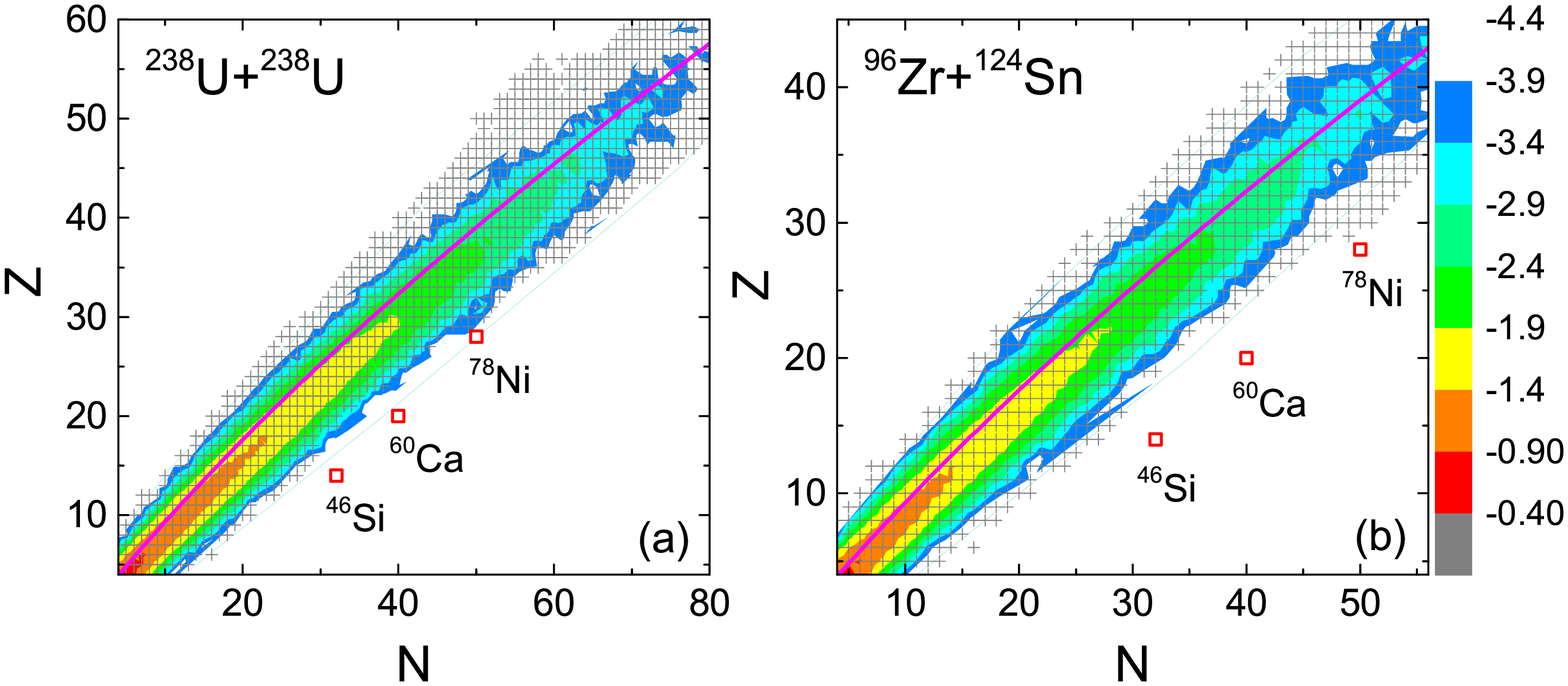}
\caption{(Color online) Predicted isotope distribution of fragments in $^{238}$U+$^{238}$U and $^{96}$Zr+$^{124}$Sn at an incident energy of 35 AMeV and $b=1$ fm (in logarithmic scale). The curves denote the $\beta$-stability line and the crosses denote the nuclei with known masses in AME2012 \cite{Audi12}.  }
\end{figure}

In Fig. 7(a), we show the calculated charge distribution of $^{129}$Xe+$^{120}$Sn at impact parameter $b=1$ fm. The curves denote the experimental data for $^{129}$Xe+$^{\rm nat}$Sn taken from \cite{Taba}. The circles denote the calculated results of ImQMD-v2.2 at incident energy of 50 AMeV. Here, the ImQMD simulations are performed till $t=2000$ fm/c. We have also checked that the charge distribution of primary fragments remains unchanged generally with a further time evolution of a few hundreds fm/c. We find that the predictions of the ImQMD model are in good agreement with the data. In Fig. 7(b), we present the isotope distribution of fragments in logarithmic scale. The solid curve denotes the positions of $\beta$-stability line described by Green's formula. We note that the number of neutron-rich fragments produced in this reaction is larger than that of proton-rich ones. To produce new neutron-rich isotopes, we also investigate the central collisions of $^{238}$U+$^{238}$U and neutron-rich system $^{96}$Zr+$^{124}$Sn. Fig. 8 shows the isotope distribution of fragments in the central collisions of $^{238}$U+$^{238}$U and $^{96}$Zr+$^{124}$Sn at an incident energy of 35 AMeV. We create 20000 events for each reaction. For each event, the microscopic dynamics process is self-consistently simulated till $t=2000$ fm/c. Fig. 8 shows the calculated isotope distribution of primary fragments. The squares present the positions of the three magic nuclei $^{46}$Si, $^{60}$Ca, $^{78}$Ni mentioned previously. From Fig. 8, some unmeasured extremely neutron-rich fragments with $Z=16-28$ can be evidently observed in U+U reaction. However, for the central collisions of neutron-rich reaction system $^{96}$Zr+$^{124}$Sn, almost all fragments produced are located in the region where the nuclear masses have already been measured. The estimated production probability for the unmeasured nuclei $^{60}$Ca and $^{78}$Ni is smaller than $5 \times 10^{-5}$ in the cental collisions of U+U at $E=35$ AMeV. According to the predicted isotope distribution of U+U at 35 AMeV, the production probability of $^{46}$Si might be much smaller than that of $^{78}$Ni.

\begin{center}
\textbf{V. SUMMARY}
\end{center}

In this work, the Fermionic properties of nucleons are investigated in the ImQMD simulations and the multi-fragmentation process in heavy-ion collisions at intermediate energies are explored for producing new neutron-rich isotopes. By introducing the momentum re-distribution via two-body "inelastic scattering" in the Fermi constraint procedure, the stability of initial nuclei and fragments is further improved. The number of "pseudo" nucleons (with $\bar f_{i}>1$) is effectively suppressed with the modified procedure in the Fermi constraint. Simultaneously, the energy conservation of the system is obviously improved in the heavy-ion reactions and in the individual nuclei. With the revised version ImQMD-v2.2 of the model, the fusion excitation functions of $^{16}$O+$^{186}$W and the fragmentation in the central collisions of Au+Au at 35 AMeV are re-examined, and the fusion cross sections at sub-barrier energies and the charge distribution of fragments are relatively better reproduced. In addition, the charge and isotope distribution of fragments in Xe+Sn, U+U and Zr+Sn at intermediate incident energies are systematically investigated with the ImQMD-v2.2 model. Some extremely neutron-rich fragments with $Z=16-28$ are evidently observed in the central collisions of $^{238}$U+$^{238}$U at 35 AMeV, whereas in neutron-rich reaction system $^{96}$Zr+$^{124}$Sn, almost all fragments produced are located in the region with known masses.

\begin{center}
\textbf{ACKNOWLEDGEMENTS}
\end{center}
We thank Dr. Yingxun Zhang for a reading of the manuscript and helpful discussions. This work was supported by National Natural Science Foundation of
China (Nos. 11275052, 11265004, 11365004, 11365005, 11422548, 11547307) and the Open Project Program of State Key Laboratory of Theoretical Physics, Institute of Theoretical Physics, Chinese Academy of Sciences, China (No. Y4KF041CJ1). The code of the ImQMD model is available at
http://www.imqmd.com/code/.

\end{document}